\documentstyle[12pt]{article}
\newcommand{\be}{\begin{equation}}

\newcommand{\ee}{\end{equation}}

\newcommand{\ba}{\begin{eqnarray}}
\newcommand{\ea}{\end{eqnarray}}
\newcommand{\baa}{\begin{eqnarray*}}
\newcommand{\eaa}{\end{eqnarray*}}

\newcommand{\eb}{\end{thebibliography}}

\newcommand{\ci}[1]{\cite{#1}}

\newcommand{\bi}[1]{\bibitem{#1}}

\begin{document}

\bigskip\bigskip

\begin{center}
{\Large \bf GERASIMOV-DRELL-HEARN AND BURKHARDT-COTTINGHAM SUM RULES
FOR PROTON, NEUTRON AND DEUTERON}
\\[1cm]
{O.V. Teryaev}\\[0.3cm]
{\it Bogoliubov Laboratory of Theoretical Physics\\
Joint Institute for Nuclear Research, Dubna\\
141980, Russia}
\footnote{E--mail: teryaev@thsun1.jinr.ru} \\[1.2cm]
%\end{center}

\begin{minipage}{150mm}
\begin{abstract}
%\centerline{\bf Abstract}

A description of the generalized Gerasimov-Drell-Hearn
sum rules for proton, neutron and deuteron is suggested.
The proton consideration confirms the earlier conjecture, that the
structure function $g_T$ features a smooth $Q^2-$dependence,
while the structure function $g_2$ is changing rapidly,
due to the elastic contribution to the Burkhardt-Cottingham sum rule.
The neutron case features the close connection to the Bjorken sum rule.
The deuteron case reveals the crucial role of photodesintegration channel,
resulting in the strong nuclear correction at very low $Q^2$.
The new approximate sum rules for this channel
reflecting the delicate interplay between nucleon and nuclear structure
are suggested.
\end{abstract}

\end{minipage}
\end{center}

%\newpage
\section{Introduction}

The generalized ($Q^2$-dependent) Gerasimov-Drell-Hearn (GDH) sum rules
provide the important link between the low energy description of hadrons
by classical GDH \cite{Ger,DH} sum rules for real photons ($Q^2=0$)
and PQCD description at large $Q^2$.
They are
just being tested experimentally with a high accuracy for both proton and
neutron \cite{E143,JLAB,HERMES}.
The striking feature of the proton data is the low ($\sim 200-250 MeV^2$)
``crossover'' point,  which is in complete agreement
with the prediction \cite{ST93,ST95}, published almost 10 years ago.
It was based on the relation to another fundamental sum rule,
namely, Burkhardt-Cottingham (BC)
sum rule for structure function $g_2$. Its elastic contribution
is the main source of strong $Q^2-$dependence, while
the contribution of the other structure function, $g_T=g_1+g_2$ is smooth.
In the present report I review this approach for proton,
as well as more recent treatment of neutron using the connection
to Bjorken sum rule.
Also, following the general line of this conference, exploring the nuclear
aspects of QCD, I will address the problem of (generalized) GDH
and BC sum rule for deuteron. I will use the approximation of zero
binding energy and show, that the crucial role in the description
of nuclear effects is played by the
photodesintegration channel. Its effect is manifested at rather low $Q$
of the order of pion mass and leads to the set of the new approximate
"finite-energy" sum rules for photodesintegration channel.
%-----------------------------------------------
%-----------------------------------------

\section{Proton: Burkhardt-Cottingham sum rule for longitudinally
polarized particles \cite{ST93,ST95}}

Let us consider the $Q^2$-dependent integral
\be
I_1(Q^2)={2 M^2\over {Q^2}} \int^1_0 g_1(x,Q^2) dx.
\label{I1}
\ee
It is defined for {\it all} $Q^2$, leaving no ambiguity
for the choice of "virtual photon flux factor", and $g_1$ is the obvious
generalization for all $Q^2$ of the standard scale--invariant $g_1(x)$.
Therefore, the comparison with its another definitions should be made
with some care, although the ambiguity due to the choice of
smooth function $I_T$ (see below) makes this discrepancy
actually beyond the accuracy of our approach.
Note that the elastic contribution at $x=1$ is not
included in the above sum rule. One recovers then at $Q^2=0$
the GDH sum rule
\be
\label{gdh0}
I_1(0)=-{\mu_A^2 \over 4}
\ee
where $\mu_A$ is the nucleon anomalous magnetic moment in nuclear magnetons.
While $I_1(0)$ is always negative, its value at large $Q^2$ is determined
by the $Q^2$ independent integral $\int^1_0 g_1(x) dx$, which is
positive for the proton and negative for the neutron.

The separation of the contributions of $g_T$ and $g_2$
leads to the decomposition of  $I_1$ as the
difference of the contributions
of two form factors $I_{T}$ and $I_2$
\be
I_1=I_{T}-I_2,
\ee
where
\be
I_{T}(Q^2)={2 M^2\over {Q^2}} \int^1_0 g_{T}(x,Q^2) dx,
\;\;\;\;I_2(Q^2)={2 M^2\over {Q^2}} \int^1_0 g_2(x,Q^2) dx;
\ee
\be
g_{T}=g_1 + g_2.
\ee
There are solid theoretical arguments to expect a strong $Q^2$-dependence
of $I_2$. It is the well-known Burkhardt-Cottingham sum rule \ci{BC},
derived independently by Schwinger \ci{Sch},
using a rather different method. It states that
\be
\label{el}
I_2(Q^2)={1\over 4}\mu G_M (Q^2)
\frac{\mu G_M (Q^2) - G_E (Q^2)}{1+\frac{Q^2}{4M^2}},
\ee
where $\mu$ is the nucleon magnetic moment, $G$'s denoting the familiar
Sachs form factors which are dimensionless and normalized to unity
at $Q^2=0$. For large $Q^2$ one can neglect the r.h.s.
of (\ref{el}) and get
\be
\int^1_0 g_2(x) dx=0.
\ee

In particular, from Eq.(7) it follows that
\be
I_2(0)={\mu_A^2+\mu_A e \over 4},
\ee
$e$ being
the nucleon charge in elementary units.
To reproduce the GDH value (see Eq.(3)) one should have
\be
I_{T}(0)={\mu_A e \over 4},
\ee
which was indeed proved by Schwinger \cite{Sch}.
The importance of the $g_2$ contribution can already
be seen, as the entire $\mu_A-$ term for the GDH sum rule is
provided by $I_2$.

Note that $I_{T}$ does not differ from $I_1$ for large $Q^2$ due to the
BC sum rule, but it is {\it positive} in the proton case.
It is possible
to obtain a smooth interpolation for  $I^p_{T}(Q^2)$
between large $Q^2$ and $Q^2=0$ \ci{ST93}.

\be
I^p_{T}(Q^2)=\theta(Q^2_0-Q^2)({\mu_A \over 4}- {2 M^2 Q^2\over
{(Q^2_0)^2}} \Gamma^p_1)+\theta(Q^2-Q^2_0) {2 M^2\over {Q^2}}
\Gamma^p_1,
\ee
where $\Gamma^p_1=\int^1_0 g^p_1(x) dx$.
The continuity of the function and of its derivative is guaranteed
with the choice $Q^2_0=(16M^2/\mu_A) \Gamma^p_1 \sim 1GeV^2$,
where the integral is given by the world average proton data.
It is quite reasonable to distinguish
the perturbative and the non-perturbative regions. As a
result one obtains a crossing point at $Q^2 \sim 0.2 GeV^2$, below the
resonance region \ci{ST93}, while the asymptotic behaviour for $Q^2 > 1 GeV^2$ and 
positive value at $Q^2=0.5 GeV^2$
is in a good agreement with the HERMES \cite{HERMES},  
E143 \ci{E143} and JLAB \cite{JLAB} data.
This smooth interpolation seems to be very reasonable in the
framework of the QCD sum rules method as well. Then one should choose  some
"dominant" tensor structure to study the $Q^2$-dependence of its scalar
coefficient and $T_1$ appears to be a good candidate. This seems
also promising  from another point of view. It is not trivial to
obtain within the QCD sum rules approach the  GDH value at $Q^2=0$.
Since the r.h.s. of (12) is linear in $\mu_A$, it may be possible to
obtain it using the Ward identities, just like the normalization
condition for the pion form factor. This, in turn opens
the possibility to apply the powerful tool of quark-hadron duality.
The latter relies on the perturbative theory, and it is quite clear
that it is much more plausible to describe linear (one-loop), than quadratic
terms. One should recall here that while sum rule for $I_{T}$ was checked
in QED long ago \ci{Mil}, the GDH sum rule required much more efforts
\ci{GDHQED}, as it gets non-trivial contributions only at two loops order,
although zero appearing at one loop level is also non-trivial \ci{Alt,BS}.

Note also that large contribution of $g_2$ by no mean contradicts
the resonance approaches \ci{Ioffe} and may be
considered complimentary to them.
The central
role here plays the $\Delta(1232)$: it provides a significant amount
of GDH integral at $Q^2=0$ and gives a clear qualitative explanation
of rapid $Q^2$-dependence\ci{book}. The $\Delta$ photoproduction is
dominated by the magnetic dipole form factor, leading to a negative
$I_1$. The sign change is just related  to the fast decrease of the
$\Delta$ contribution.
In order to compare this picture with our approach we have
separated \ci{ST95}
the $\Delta$ contribution to $I_{T}$ and $I_2$. To do this we
just calculated the photoproduction Born diagram using the well-known
expressions for the covariant form factors $G_M$, $G_E$ and $G_C$. The
resulting expression is rather lengthy but it has a remarkable
property: the leading $G_M^2$ term contains {\it only} the tensor
$T_2$ at any $Q^2$. This fact is confirmed if one performs the
contraction with the virtual photon density matrix: in particular, if
one takes the standard definition with the kinematical zero at
$Q^2=0$, the result should be attributed to $g_1$. In approach \cite{ST95}
the nonzero $g_1$ is due to the absence of $T_1$ since $g_1=-g_2$.

\section{Neutron: link to Bjorken sum rule \cite{ST02}}

To generalize this approach to the neutron case, one needs a
similar smooth parameterization of $g_T$
for the neutron. Since the value at
$Q^2=0$ is equal to zero, it is not sufficient to limit oneself
to the simplest linear parameterization. One needs to
add a term, quadratic in $Q^2$. A simple parameterization
providing the continuity of the function and its derivative
was suggested in \ci{ST97}, which however, leads to the result
being in contradiction with the data.
This does not seem to be occasional, bearing in mind the
argument presented above. Indeed, the general reason
supporting the smoothness of interpolation for $g_T$ is
its linearity in $\mu_A$. As soon as this term appears to be equal to
zero for some occasional reason (which, in our case, is nothing else than
the neutrality of neutron!), there is also no more reason to expect the
smoothness as well.

To bypass this obstacle, we use the difference between the proton and
neutron instead of neutron itself. Although it is possible, in principle,
to construct the smooth interpolation for functions $g_1$ themselves
\ci{book}, it also does not fit the suggested general argument, as
r.h.s. is proportional to $\mu_{A,p}^2- \mu_{A,n}^2$, so that
it is quadratic and, moreover,
has an additional suppression due to the smallness
of isoscalar anomalous magnetic moment. Indeed, the smooth approximation
for $g_1$ resulted in the $Q^2 \sim 5 GeV^2$, leading to the rather early
scaling violations, do not observed experimentally.
Insteadt, it was recently suggested
the following parameterization for isovector
formfactor $I^{p-n}_{T}$:

\be
I^{p-n}_{T}(Q^2)=\theta(Q^2_1-Q^2)({\mu_A \over 4}- {2 M^2 Q^2\over
{(Q^2_1)^2}} \Gamma^{p-n}_1)+\theta(Q^2-Q^2_1) {2 M^2\over {Q^2}}
\Gamma^{p-n}_1,
\ee
where transition value
$Q_1^2$ may be determined by the continuity
conditions in a similar way,
and we get value $Q_1^2 \sim 1.3 GeV^2$, being of the same order
as in the proton case.

The elastic contribution to BC sum rule should be included
for neutron separately, so
we need
the neutron elastic form factors. While the electric one, might be
neglected, the magnetic form factor
is well described by the dipole formula \cite{FF}

\be
G_M (Q^2)={1 \over {(1+Q^2/0.71)^2}}.
\ee
%The plots representing $\Gamma^n_1 (Q^2)$ are displayed
%on Fig.1.
%\begin{figure}[t]
%\begin{minipage}[t]{75mm}
%\centerline{\epsfbox{fign.ps}}
%\centerline{\epsfxsize15.0cm\epsffile{fign.eps}}
%\centerline{\includegraphics[scale=0.6]{fign.eps}}
%\caption{{\protect\small Neutron structure function $g_1$}}
%at
%$m_{2\pi}=0.98$
% GeV$^2$, $Q^2=5$ GeV$^2$ }}
%\label{plotphase}
%\end{minipage}
%\end{figure}
As a result \cite{ST02}, the neutron $g_1$
stays quite close to its asymptotic value
until rather low $Q$ are achieved. Moreover, the experimental data happen
to bear some traces of the structure, generated in our approach
by the interplay of $Q_0$ and $Q_1$.

\section{Deuteron: sum of squares vs square of sum and new finite energy
sum rules for photodesintegration channel}

The deuteron case poses the new problem, which is manifested
already for real photon scattering ($Q^2=0$).
Considering deuteron as a
fundamental particle of spin 1 one get the deuteron
anomalous magnetic moment, which is very small due to the approximate
cancellation of proton and neutron AMM,
in the r.h.s. of (\ref{gdh0}).

At the same time, for large energies (and/or $Q^2$ ) one
should expect, due to the small binding energy
to get the incoherent sum of the proton and neutron
contributions which is rather large.

To make this effect even more dramatic one may consider the following
{\it gedankenexperiment}. Let the binding to be adibatically switched on.
Than, for zero binding, one should get the {\it sum of squares}
of the proton and neutron AMM, while for {\it arbitrary} weak binding
it should become equal to the {\it square of their sum},
as weakness of binding will make the deviation from AMM additivity negligible.

So, what transforms the sum of squares to square os sums?
The answer is almost obvious. For arbitrary small binding there
is a new channel, photodesintegration $d \to p+n$, absent for zero binding,
when it corresponds to elastic contribution do not entering to GDH integral.

As soon as binding energy is small, one may assume that the
inelastic contributions (starting from the pion threshold)
for free proton and neutron are not affected by binding. If so,
the inelastic contribution to deuteron GDH integral
is equal to the sum of the ones for free proton and neutron.
As a result, one may get the following finite energy sum rule:

\be
\label{fesr}
\int_{\nu_{PD}}^{\nu_\pi} \frac {d \nu}{\nu} \Delta \sigma_d(\nu)=
-\frac{4 \pi^2 \alpha}{M^2} ((\sum \mu_A)^2-\sum \mu_A^2) \approx
-\frac{8 \pi^2}{M^2} \mu_A^P \mu_A^n
%\frac{2 \pi^2}{M^2}\mu_A^P \mu_A^n
\ee

Here $\nu_{PD}$ and $\nu_\pi$
are the photodesintegration and inelastic (one-pion)
thresholds, respectively.
The accuracy of this sum rule is guaranteed by smallness of deuteron
binding energy $\epsilon$. The relevant small parameter may be estimated
as its ratio $\beta$ to the minimal energy, which should be
transferred to nucleon
to produce the pion,
\be
\label{al}
\beta =\frac {2 M \epsilon}{ m_\pi^2} \sim 10\%.
\ee
The contribution of photodesintegration channel was first estimated
long ago by S.B. Gerasimov \cite{ger2},
but the use of non-relativistic approximation
resulted in the correction to (\ref{fesr}) of about $50\%$, so that the
significant role of nuclear effects at high-energies was expected.
However, the more elaborate recent relativistic calculation
\cite{aren} led to the validity
of (\ref{fesr}) with the accuracy of about $10\%$, which is the one expected
from the small binding energy approximation (SBEA)suggested above.

This approximation should become much more accurate in the virtual photon case,
as now
$$\beta =\frac {2 M \epsilon}{ m_\pi^2+Q^2}.$$
This mean that it should provide very good description down to $Q^2 \sim m_\pi^2$
(recall, that even for $Q^2 \sim 0$ the accuracy of SBEA is still rather good,
so that it is $m_\pi^2$ which determines the characteristic value of $Q^2$).

Consequently, down to low $Q^2 \sim m_\pi^2$ the GDH integral should be very
close to the sum of the proton and neutron values. The difference between
"square of sum" and "sum of squares" (which, in turn is rapidly
decreasing because of the decrease of the form factors) should be provided
by the photodesintegration channel contribution, corresponding to
very large  $x_B \sim 1 - 2 M m_\pi/Q^2$ and therefore also small.

So, it is convenient to decompose the generalized GDH SR and consider the
photodesintegration and inelastic channels separately. While the inelastic
channel contribution should be close to the sum of proton and neutron down to
real photon limit, while the photodesintegration one should rapidly grow at
low $Q^2$ and tend to the value (\ref{fesr}).

One can immediately make more quantitative predictions for BC SR,
valid for all $Q^2$. Like in the case of generalized GDH SR, SBEA
should be good at all $Q^2$, resulting in another sum rule for
photodesintegration
channel.

\be
\label{febc}
%\int_{\nu_{PD}}^{\nu_\pi} d\nu G_2(\nu) =
\int_{x_\pi}^{x_{PD}} g_2^d (x) dx
=\frac{1}{4} ((\sum \mu G_M)^2 - \sum \mu G_M \sum G_E - \sum
\mu G_M(\mu G_M - G_E)),
%-\frac{4 \pi^2}{M^2} \mu_A^P \mu_A^n
\ee
where $ x_\pi, x_{PD}$ correspond to inelastic and photodesintegration
channels, respectively.
This sum rule may be probably checked at JLAB,
while (\ref{fesr}) may be studied
at TUNL \ci{weller}.

I am indebted to J. Soffer for the fruitful collaboration in this field and to
S.B. Gerasimov and A. Sch\"afer for valuable comments. This investigation was supported
by INTAS(Contract 00-00587)

\begin{thebibliography}{99}
\bi{Ger} S. B. Gerasimov, Yad. Fiz. {\bf 2}, 598(1965)
[Sov. J. Nucl Phys. {\bf 2}, 430(1966)].
\bi{DH} S. D. Drell and A. C. Hearn, Phys. Rev. Lett. {\bf 16}, 908 (1966).
\bi{E143} E143Collaboration, K. Abe et al., Phys. Rev. Lett.{\bf78}, 815
(1997).
\bi{JLAB}M.~Amarian {\it et al.},
arXiv:nucl-ex/0205020.

\bi{HERMES} 
A.~Airapetian {\it et al.}  [HERMES Collaboration],
%``The Q**2-dependence of the generalised Gerasimov-Drell-Hearn integral  for the deuteron, proton and neutron,''
Eur.\ Phys.\ J.\ C {\bf 26}, 527 (2003)

\bi{ST93} J. Soffer and O. Teryaev, Phys. Rev. Lett. {\bf 70}, 3373 (1993).
\bi{ST95} J. Soffer and O. Teryaev, Phys. Rev. {\bf D51}, 25 (1995).
\bi{BC} H. Burkhardt and W. N. Cottingham, Ann. Phys. (N.Y.)
{\bf 16}, 543 (1970).
\bi{Sch} J. Schwinger, Proc. Nat. Acad. Sci. U.S.A. {\bf 72}, 1559 (1975).
\bi{Mil}
Wu-Yang Tsai, L.DeRaad and K.A.Milton, Phys. Rev. {\bf D11},
3537(1975).
\bi{GDHQED} D.A. Dicus and R. Vega, Phys. Lett. {\bf B501}, 44 (2001)
\bi{Alt} S.B. Gerasimov and J. Moulin, JINR E2-6722, 1972.
G. Altarelli, N. Cabibbo and L. Maiani,
Phys. Lett. {\bf B40}, 415 (1972)
\bi{BS} S.J. Brodsky and I. Schmidt, Phys. Lett. {\bf B351}, 344 (1995)
\bi{Ioffe}
V. D. Burkert and B. L. Ioffe, Phys. Lett. {\bf B296}, 223 (1992),
J.\ Exp.\ Theor.\ Phys.\  {\bf 78}, 619 (1994)
[Zh.\ Eksp.\ Teor.\ Fiz.\  {\bf 105}, 1153 (1994)];
V.~D.~Burkert, arXiv:nucl-ex/0109004.
\bi{book} B. L. Ioffe, V. A. Khoze, L. N. Lipatov, {\it Hard Processes},
(North-Holland,Amsterdam, 1984).

\bi{ST02} J.~Soffer and O.~V.~Teryaev,
%``Bjorken sum rule at low Q**2,''
Phys.\ Lett.\ B {\bf 545}, 323 (2002)
\bi{ST97} J. Soffer and O. Teryaev, Phys. Rev. {\bf D56}, 7458 (1997).

\bi{ger2} S.B. Gerasimov, Yad. Fiz. {\bf 5}, 1263 (1967)
\bi{aren} H. Arenh\"ovel, nucl-th/0006083
%\bi{NNN}
%I.~P.~Ivanov et al.,
%, N.~N.~Nikolaev, A.~V.~Pronyaev and W.~Schafer,
%Phys.\ Lett.\ B {\bf 457}, 218 (1999)
%\bi{FF}  A. Lung et al.,Phys. Rev. Lett. {\bf 70}, 718 (1993).
\bi{weller}
H.R. Weller,
Published in *Mainz 2000, Gerasimov-Drell-Hearn sum rule and the nucleon spin
structure in the resonance region* 145-156

\eb

\end{document}